\documentclass[aps,prd,amsmath,amssymb,12pt]{revtex4}
\usepackage{graphicx}
\usepackage{bm}
\def\journal#1#2#3#4{{#1} {\bf #2}, #3 (#4)}
\newcommand{\be}{\begin{equation}}
\newcommand{\ee}{\end{equation}}
\newcommand{\bea}{\begin{eqnarray}}
\newcommand{\eea}{\end{eqnarray}}
\newcommand{\hf}{\frac12}
\newcommand{\nn}{\nonumber\\}
\def\eq#1{(\ref{#1})}
\def\la{\langle}
\def\ra{\rangle}

\def\Tr{{\mathrm{Tr}}}
\def\ord#1{{\cal O}\left(#1\right)}
\def\mr#1{{\mathrm{#1}}}
\def\v#1{{\bm{#1}}}
\def\fd#1#2{\frac{\delta#1}{\delta#2}}
\def\fdd#1#2#3{\frac{\delta^2#1}{\delta#2\delta#3}}
\def\dt{\Delta t}
\def\dv#1{\dot{\v{#1}}}
\def\ddv#1{\ddot{\v{#1}}}

\def\hj{{\hat j}}
\def\hpsi{{\hat\psi}}
\def\hpsid{{\hat\psi^\dagger}}
\def\hs{\hat\sigma}

\def\hx{\hat x}
\def\hy{\hat y}
\def\hD{{\hat D}}
\def\hG{{\hat G}}
\def\ih{\frac{i}{\hbar}}

\def\psid{\psi^\dagger}
\def\sign{\mr{sign}}

\begin{document}
\title{Dissipation and decoherence by a homogeneous ideal gas}
\author{Janos Polonyi}
\email{polonyi@iphc.cnrs.fr}
\affiliation{Strasbourg University, CNRS-IPHC, 23 rue du Loess, BP28 67037 Strasbourg Cedex 2, France}

\begin{abstract}
The effective Lagrangian of a test particle, interacting  within an ideal gas, is calculated within the closed time path formalism in the one-loop approximation and in the leading order of the particle trajectory. The expansion in the time derivative, available for slow enough motion, uncovers diffusive forces and decoherence in the particle coordinate basis. The master equation, generated by the effective Lagrangian, is derived and its consistency is verified for a finite-temperature gas.
\end{abstract}
\maketitle

\section{Introduction}
The time-reversal invariance of effective forces is a nontrivial issue: Even if the dynamics of a closed system is time-reversal invariant, the effective interaction between a subsystem and the rest, its environment, always breaks the time-reversal invariance due to the environment boundary conditions in time. A further layer of complications is found in infinite systems where the effective forces may be dissipative. Quantum systems raise an additional question about the decoherence, another irreversible process, generated by the effective interactions. These issues had already attracted much attention, and the emergence of irreversibility has been demonstrated in a physically appealing way, by coarse graining \cite{zwanzig}. In a similar manner, decoherence can be generated by the large, highly degenerate environment \cite{zeh,zurek}. Such a general picture naturally leaves open the details about the actual loss of information, realized by the use of a restricted set of observables, in a given microscopic model.

The traditional approach to Brownian motion comes from kinetic theory; the quantum theory is inferred from a master equation. The master equation, obtained first for a particle colliding with a gas \cite{zeh}, was shown to describe the decoherence while dissipation has been ignored. Though the resulting divergence for long time can be eliminated without taking account of dissipation \cite{gallis}, the physically satisfactory description requires the presence of both the dephasing and the dissipation \cite{diosi,adler}. A systematic, perturbative derivation of the master equation, based on the collision cross section of the test particle, taken in the Born approximation, has been worked out, too \cite{vacchini,lanz,vacchinie}. This approach was later improved by going beyond the Born approximation \cite{hornbergerk}. The traditional many-body technique is available, as well, to arrive at a master equation \cite{dodd}.

Another way to approach this problem is model building. The simplest, exactly solvable model consists of a particle, coupled linearly to infinitely many harmonic oscillators \cite{agarwal,calderialeggett,unruh}. The effective dynamics of an open system, the test particle in the present case, can easiest be handled within the closed time path (CTP) scheme. This formalism was first introduced in quantum field theory \cite{schw} and has since then been successfully applied in different areas of condensed matter physics \cite{keldysh,kamenev,rammer} and particle physics \cite{calzetta}. The harmonic model \cite{dekker} has thoroughly been studied in this scheme \cite{grabert} and a general form of the master equation for the reduced density matrix, containing a memory term, has been derived in the CTP formalism \cite{hu9293}. 

Though the Markovian approximation is believed to be applicable at high enough temperatures, the high temperature master equation, derived in Ref. \cite{calderialeggett}, failed to preserve the positivity of the density matrix \cite{stenholm,ambegaokar,munro}. A Markovian master equation of the Lindblad form \cite{lindblad}, i.e., satisfying the physical requirements for the density matrix, has been found for high enough temperatures \cite{diosip}, fitting into the set of minimally invasive generalizations \cite{munro,breuerp} of the master equation of Ref. \cite{calderialeggett}. A physically satisfactory master equation can be constructed in an ad hoc manner \cite{gao} and by the use of an appropriate ansatz for the $T$ matrix in the collisional approach \cite{vacchini}, as well.

The goal of this paper is a systematic derivation of the effective dynamics of a test particle, moving in a gas. We shall consider an ideal gas, the only interaction in the model being between the test particle and the individual gas particles. The fact that in this model dissipative forces occur is not surprising since the leading-order, one-loop contribution to the transport coefficients which is equivalent to using Kubo's formulas \cite{kubo} can be interpreted as the contribution of an ideal gas. The effective Lagrangian of the test particle is derived within the framework of the Landau-Ginsburg double expansion, assuming that the interaction with the gas generates a small amplitude, slow modification of the test particle trajectory. The effective action of a particle which interacts with an ideal gas was calculated some time ago by using the traditional effective action approach in imaginary time \cite{guinea} and by means of the CTP formalism \cite{hedegard}. Though our procedure is similar to the one followed in these works, the final form of the effective action is different. This is because we had to go beyond the traditional action formalism to recover dissipative forces and did the separation of the conservative and the dissipative forces in an appropriate way. Another element of this work is that the decoherence of the coordinate has been monitored and the result has been compared with those of the harmonic model \cite{calderialeggett}. It is found that both the friction forces and the decoherence are generated by the same quantity in our one-loop effective theory, supporting the view about the common origin of dissipation and decoherence.

Although the collision-based and the CTP effective action descriptions have different starting points they share some common assumptions. In the collisional picture one assumes a dilute gas to truncate the hierarchy equations, it requires short range interactions to rely on scattering processes, and it needs high enough temperatures to render the effective dynamics Markovian. Finally, the Brownian limit is carried out by letting the ratio of the gas particles and test particle mass go to zero to arrive at a simpler equation of motion for the reduced density matrix. These assumptions are mirrored and partially softened in the effective action approach where the effective action is usually constructed within the framework of the Landau-Ginsburg double expansion. In fact, the expansion in the amplitude is reminiscent of the assumption of weak interactions in a dilute gas. The expansion in the gradient, the assumption that the effective interactions are local, corresponds to the use of scattering processes in the other scheme. The calculation of the higher order contributions is not exceedingly difficult in the effective action, leaving open the way to a systematic inclusion of multi-particle correlations and non-Markovian effects. Notice that the light gas particle limit is not necessary for the calculation of the effective action.

The calculation of the order $\ord{x^2}$ and $\ord{\partial_t^2}$ effective Lagrangian is presented below. The real part displays Newton's friction force and a mass renormalization whereas the imaginary part describes the decoherence of the coordinate. The result simplifies considerably when an ideal gas of fermions is considered at vanishing  temperature: Both the friction force and the strengths of decoherence stem from the same source and are proportional to the particle velocity. Furthermore, the decoherence shows a characteristic anisotropy, namely it is six time stronger in the direction of the velocity than in the perpendicular plane. The master equation for the density matrix, generated by the effective Lagrangian, is derived and its Lindblad form is verified for finite temperature fermion gas. It is remarkable that one needs the full $\ord{\partial_t^2}$ decoherence dynamics to establish this result. 

The paper starts with a short summary of the CTP formalism, presented first for a closed system in Sec. \ref{closeds} when it appears as a trivial rewriting of the traditional, transition amplitude based formalism of quantum mechanics. The Green's functions are introduced for a single harmonic oscillator in Sec. \ref{shos}. The CTP formalism becomes unavoidable for open systems when the system-environment interactions are non-conserving and make the system dynamics open. The resulting effective theory is briefly introduced in Sec. \ref{opens}. Such an  effective theory is calculated for a harmonic toy model in Section \ref{toys}. The main topics of this work, the effective dynamics of a test particle within a gas is  embarked in Sec. \ref{partingass}. It starts in Section \ref{effths} with outlining the perturbative derivation of the effective theory for the test particle. The general equations, obtained there, are used in Section \ref{quadreffs} to find the $\ord{x^2}$, $\ord{\partial_t^2}$ effective Lagrangian, the special case of an ideal gas of fermions at vanishing temperature being touched upon in Section \ref{idfgvts}. The master equation, corresponding to this Lagrangian, is derived and its Lindblad structure is verified in Section \ref{masters}. Finally, Section \ref{concls} contains our conclusions.

\section{CTP formalism}\label{ctps}
The CTP formalism has been proposed to deal with expectation values in the Heisenberg representation \cite{schw} rather than transition amplitudes of the Schr\"odinger picture. Though this is already an important extension compared to the traditional, transition amplitude based formalism the real strength of this scheme becomes visible for open systems, interacting with their environment. The effective dynamics of such systems, such as dissipation, can not be captured by relying on pure states only and the use of mixed state and the reduced density matrix is needed. This is the point where CTP formalism becomes unavoidable, being the only scheme to deal with mixed states in many-body systems and to visualize the elementary processes by borrowing the clarity of the Feynman diagrams.

\subsection{Closed system}\label{closeds}
Let us suppose that our closed quantum system is described by the coordinate $x$ and a time independent Hamiltonian, $H$. The time evolution operator, 
\be\label{ctpgfncto}
U(t_f,t_i;j)=T[e^{-\ih\int_{t_i}^{t_f}dt'[H-x(t)j(t)]}],
\ee
where $T$ denotes the time ordering and the source, $j(t)$, a book-keeping device, is written in the Heisenberg representation and allows us to recover the expectation value of an $x$-dependent observable, $A(x)$, in the form 
\be\label{expv}
\la A(x(t))\ra=\Tr[A(x(t))\rho]=\Tr[A(x)U(t,t_i;0)\rho_iU^\dagger(t,t_i;0)],
\ee
where $\rho_i$ denotes the density matrix of the initial state. The calculation of this quantity is facilitated by the use of the generator functional,
\be\label{ctpgfunc}
e^{\ih W[\hj]}=\Tr[U(t_f,t_i;j^+)\rho_iU^\dagger(t_f,t_i;-j^-)],
\ee
where the CTP doublet,  $\hj=(j^+,j^-)$, contains two independent sources, introduced for each time evolution operator in Eq. \eq{expv}, and serve to generate the observable and the interaction vertices of the perturbation series by the functional derivatives $\delta/\delta j^\pm(t)$, e.g.
\be\label{expval}
\la A(x(t))\ra=A\left(-i\hbar\fd{}{j^\sigma(t)}\right){e^{\ih W[\hj]}}_{\hj=0},
\ee
with $\sigma=\pm$. Note that the unitarity of the time evolution assures that either source can be used to generate the expectation value for arbitrary $t_f>t$. The path integral representation of the generator functional, \eq{ctpgfunc}, can be obtained by the slicing of both time evolution operators according to the strategy of the traditional path integration for transition amplitudes. The result is an integration over the pairs of trajectories, $\hx(t)=(x^+(t),x^-(t))$, 
\be\label{ctpqmgf}
e^{\ih W[\hj]}=\int D[\hx]e^{\ih S[\hx]+\ih\int dt\hj(t)\hx(t)},
\ee
where the action,
\be\label{ctpaction}
S[\hx]=\int_{t_i}^{t_f}dt[L(x^+(t),\dot x^+(t))-L(x^-(t),\dot x^-(t))]+S_\epsilon[\hx],
\ee
contains the Lagrangian of the system, $L(x,\dot x)$, and
\be
S[\hx]=i\frac\epsilon2\int_{t_i}^{t_f}dt[(x^+(x))^2+(x^-(x))^2],
\ee
which keeps the path integral convergent for large $x$. The boundary conditions in time, suppressed in Eq. \eq{ctpqmgf}, involve the integration of the initial coordinate with the weight $\rho(x^+(t_i),x^-(t_i))$ and the final condition, $x^+(t_f)=x^-(t_f)$, representing the initial state and the trace in Eq. \eq{ctpgfunc}, respectively. 

To recover a time translation invariant action we have to take the limit $t_i\to-\infty$, $t_f\to\infty$. However, this is a rather non-trivial procedure \cite{breakd} which can be avoided by (i) considering a harmonic oscillator, as discussed below, (ii) obtaining its Green's function within the operator formalism, (iii) calculating the inverse propagator, (iv) identifying it with the kernel of the harmonic action, and finally, (v) using the kinetic energy which is obtained in such a manner as the kinetic energy for the non-harmonic case. The result is
\be\label{bcact}
S_\epsilon[\hx]=\frac{i\epsilon}2\int_{-\infty}^\infty dt[(x^+(t))^2-(x^-(t))^2]+\frac\epsilon\pi P\int_{-\infty}^\infty dtdt'\frac{x^+(t)x^-(t')}{t-t'},
\ee
where $P$ denotes the principal part prescription.

A distinguished feature of the CTP formalism, the redoubling of the degrees of freedom in the path integral, follows from the double appearance of the physical state, a bra and a ket, in the expectation value, $\la\psi|A|\psi\ra$. These states move in time in opposite directions, making it possible to represent effects which are not time-reversal invariant, and are called chronons because both degrees of freedom are needed to describe a general, not necessarily time-reversal invariant  interaction and thereby to realize the flow of the time as we imagine it \cite{breakd}. The internal time arrows of the two chronons are opposite of each others thus the chronon exchange,  $(x^+,x^-)\to\tau(x^+,x^-)=(x^-,x^+)$, represents the time-reversal transformation and the  relation,
\be\label{ctpsym}
W[\hj]=-W^*[\tau\hj],
\ee
expresses the time-reversal symmetry of the CTP formalism \cite{aharonov}.

\subsection{A single harmonic oscillator}\label{shos}
The Green's functions play an important role in solving models and they can easiest be introduced for a harmonic oscillator. We consider for this end a harmonic oscillator, defined by the Lagrangian $L_0=m_0\dot x^2/2-m\omega_0^2x^2/2$. Its action can be written in the form
\be\label{freectpact}
S_0[\hx]=\hf\int_{-\infty}^\infty dtdt'\hx(t)\hD^{-1}(t-t')\hx(t'),
\ee
and the corresponding generator functional,
\be\label{freegenf}
W_0[\hj]=-\hf\int_{-\infty}^\infty dtdt'\hj(t)\hD(t-t')\hj(t')
\ee
shows that 
\be\label{ctpgrfc}
\fdd{\ih W[\hj]}{\ih\hj(t)}{\ih\hj(t')}
=\begin{pmatrix}\la0|T[x(t)x(t')]|0\ra&\la0|x(t')x(t)|0\ra\cr\la0|x(t)x(t')|0\ra&\la0|T[x(t')x(t)]|0\ra^*\end{pmatrix}
=i\hbar\hD(t-t'),
\ee
can be interpreted as the CTP Green's function, with the Feynman propagator in its diagonal blocks and the Wightman function in the off-diagonal elements. The Green's function is defined in the frequency space by the Fourier integral
\be
\hD_\omega=\int dte^{i\omega t}\hD(t).
\ee
Eq. \eq{ctpgrfc} yields $\hD_\omega=\hG_\omega(\omega_0)/m_0$, with
\be\label{ftrgf}
\hG_\omega(\Omega)=\begin{pmatrix}\frac1{\omega^2-\Omega^2+i\epsilon}&-2\pi i\delta(\omega^2-\Omega^2)\Theta(-\omega)\cr-2\pi i\delta(\omega^2-\Omega^2)\Theta(\omega)&-\frac1{\omega^2-\Omega^2-i\epsilon}\end{pmatrix}
\ee
in a straightforward manner. The inverse of the Green's function gives, after inserting it into Eq. \eq{freectpact}, the action \eq{ctpaction} in the limit $t_i\to-\infty$, $t_f\to\infty$ with \eq{bcact}. 

The initial state in the calculation of the Green's function \eq{ftrgf} is the ground state. If the initial state contains excitation then the Green's function is modified and the calculation, leading to Eq. \eq{ftrgf}, is to be repeated. For an important family of excited initial mixed states, corresponding to thermal equilibrium one finds
\be\label{ctpscprn}
\hG_\omega(\Omega)\to\hG_\omega(\Omega)-i2\pi\delta(\omega^2-\Omega^2)n(\omega)\begin{pmatrix}1&1\cr1&1\end{pmatrix}
\ee
where $n(\omega)$ denotes the occupation number. 

The time-reversal symmetry, \eq{ctpsym}, imposes the relation $D^{\sigma,\sigma'}(\omega)=(D^{-\sigma,-\sigma'}(-\omega))^*$, which implies the characteristic CTP block structure,
\be\label{blockge}
\hD=\begin{pmatrix}D^n+iD^i_1&-D^f+iD^i_2\cr D^f+iD^i_2&-D^n+iD^i_1\end{pmatrix},
\ee
for the Green's function of any local operator in time in terms of four real functions of the time. The expectation value of the coordinate, belonging to a physical source, $j^\pm=\pm j$,
\be\label{coordev}
\la x(t)\ra=\sum_{\sigma'}\sigma'\int dt'D^{\sigma\sigma'}(t-t')j(t'),
\ee
is independent of the choice of $\sigma$, according to Eq. \eq{expval} and the Green's function of operators, obeying linear equation of motion simplifies to
\be\label{blockg}
\hD=\begin{pmatrix}D^n+iD^i&-D^f+iD^i\cr D^f+iD^i&-D^n+iD^i\end{pmatrix},
\ee
due to the $\sigma$-independence of the expectation value \eq{coordev}. Due to the symmetry, $D^{\sigma,\sigma'}(\omega)=D^{\sigma',\sigma}(-\omega)$, we have $D^n(\omega)=D^n(-\omega)$, $D^f(\omega)=-D^n(-\omega)$ and $D^i(\omega)=D^i(-\omega)$. The combination of the Green's function blocks, appearing in Eq. \eq{coordev}, $\sum_{\sigma'}\sigma'D^{\sigma\sigma'}$ gives the retarded Green's function whose even and odd components, $G^n$ and $G^f$, are traditionally called the near and the far Green's functions in electrodynamics. The positive norm of the states, contributing to the spectral function,
\be\label{spfnct}
iD^{-+}_\omega=\sum_n\la0|x_{-\omega}|n\ra\la n|x_\omega|0\ra,
\ee
and the positive excitation energies make $D^{-+}_\omega=0$ for $\omega\le0$ and $D^{-+}(\omega)\ge0$ if $\omega>0$ and induce the relation
\be\label{posencond}
iD^i_\omega=\mr{sign}(\omega)D^f_\omega.
\ee

The generator functional of a harmonic oscillator is actually given by Eq. \eq{freegenf} up to a source independent constant, corresponding to the path integral with vanishing external source. Due to the unitarity of the time evolution the generating functional  $W_0[\hj]$, defined by \eq{ctpgfunc} is vanishing for physically realizable sources, $j^+=-j^-$, $W_0[j,-j]=0$. The block structure \eq{blockg} makes $\hj\hD\hj=0$ in this case and cancels the source independent constant in $W_0[\hj]$.

\subsection{Open system}\label{opens}
The generator functional, given by Eq. \eq{ctpgfunc} can be factorized into the product of the transition amplitude and its complex conjugate for a closed system which is in a pure state. Thus the CTP scheme has no advantage over the traditional formalism in this case. But this changes when an open system is considered. Let us suppose that we observe the coordinate $x$ which interacts with its environment, described by the coordinate $y$, and the dynamics of the full, closed system is described by the action $S[x,y]=S_s[x]+S_e[x,y]$. The generator functional \eq{ctpgfunc},
\be\label{genfeffth}
e^{\ih W[\hj]}=\int D[\hx]D[\hy]e^{\ih S[x^+,y^+]-\ih S[x^-,y^-]+\ih S_\epsilon[\hx]+\ih S_\epsilon[\hy]+\ih\int dt\hj(t)\hx(t)},
\ee
assumes the form
\be\label{ctpeffgf}
e^{\ih W[\hj]}=\int D[\hat x]e^{\ih S_{eff}[\hat x]+\ih S_\epsilon[\hx]+\ih\int dt\hx(t)\hj(t)},
\ee
with $S_{eff}[\hx]=S_s[x^+]-S_s[x^-]+S_{infl}[\hx]$, where the influence functional \cite{feynman} is given by
\be\label{qinfl}
e^{\ih S_{infl}[\hx]}=\int D[\hy]e^{\ih S_e[x^+,y^+]-\ih S_e[x^-,y^-]+\ih S_\epsilon[\hy]}.
\ee

One finds a physically better motivated form of the effective action by developing the CTP formalism for the density matrix itself,
\be
\rho(x^+_f,x^-_f)=\la x^+_f|U(t_f,t_i)\rho U^\dagger(t_f,t_i)|x^-_f\ra,
\ee
rather than for its trace, c.f. Eq. \eq{ctpgfunc}. This is the open time path scheme which is based on  the generator functional for the reduced density matrix,
\be\label{effgen}
e^{\ih W[\hj]}=\sum_n\la x^+_f|\otimes\la n|U(t_f,t_i;j^+)\rho U^\dagger(t_f,t_i;-j^-)|x^-_f\ra\otimes|n\ra,
\ee
where the sum is over an environment basis. The path integral expression,
\be
\rho(x^+_f,x^-_f)=\int D[\hx]D[\hy]e^{\ih S[x^+,y^+]-\ih S[x^-,y^-]+\ih S_\epsilon[\hx]+\ih S_\epsilon[\hy]},
\ee
contains integration over chronon trajectories with final conditions $\hx(t_f)=\hx_f$ and $y^+(t_f)=y^-(t_f)$. The reduced density matrix is given by an effective path integral,
\be\label{otpr}
\rho(x^+_f,x^-_f)=\int D[\hx]e^{\ih S_{eff}[\hx]+\ih S_\epsilon[\hx]},
\ee
and it is important to bear in mind that such an effective open time path dynamics contains the effective action of the CTP formalism. 

To separate the dynamically different terms in $S_{eff}$ we use the decomposition,
\be\label{effactsep}
S_{eff}[\hx]=S_1[x^+]-S_1^*[x^-]+S_2[\hx],
\ee
with $\delta^2S_2[\hx]/\delta x^+\delta x^-\ne0$. Here the action $S_1$ describes a closed, conservative dynamics. The contributions to $S_2$ arise from the presence of several contribution to the sum over the environment basis elements in Eq. \eq{effgen}. Thus the couplings between $x^+$ and $x^-$ in $S_2$ represent the mixed state contributions to the path integral \eq{otpr}, arising from the system-environment entanglement and make the effective system dynamics open and nonconservative. 

The imaginary part of the effective action describes two different phenomena. $\Im S_2$ controls the suppression or the enhancement of the contributions to the density matrix and is a function of the separation of the chronon trajectories. In particular, if $\Im S_2[x^+,x^-]\to\infty$ as $|x^+_f-x^-_f|\to\infty$ then the density matrix is small for large $|x^+_f-x^-_f|$ and the coordinate is strongly decohered. The elementary excitations of the effective dynamics are defined by $S_1$. Their life-time is finite if $\Im S_1>0$; however, the unitarity of the time evolution is preserved by the mixed terms of the density matrix, owing to $S_2\ne0$.

\section{Toy model}\label{toys}
To prepare the calculation of the effective theory for a test particle in a gas we consider first a simpler problem within a system of linearly coupled harmonic oscillators \cite{calderialeggett}, defined by the action $S=S_s+S_e$,
\bea\label{holagra}
S_s&=&\int dt\left(\frac{m}{2}\dot x^2-\frac{m\omega^2_0}2x^2\right),\nn
S_e&=&\sum_n\int dt\left(\frac{m}2\dot y^2_n-\frac{m\omega_n^2}2y^2_n-g_nxy_n\right).
\eea
It is more advantageous to separate the system and environment in a way that the system potential reflects clearer the effective dynamics for small $\dot x$,
\bea\label{holagrc}
S_s&=&\int dt\left[\frac{m}{2}\dot x^2-\left(\frac{m\omega_0^2}2-\sum_n\frac{g_n^2}{2m\omega^2_n}\right)x^2\right],\nn
S_e&=&\sum_n\int dt\left\{\left[\frac{m}2\dot y^2_n-\frac{m\omega_n^2}2\left(y_n+\frac{g_nx}{m\omega_n^2}\right)^2\right]\right\}.
\eea
The simplest parametrization of the model is given in terms of the spectral function,
\be
\rho(\Omega)=\sum_n\frac{g_n^2}{2m\omega_n}\delta(\omega_n-\Omega).
\ee

\subsection{Effective dynamics}
The usual way of finding the effective system dynamics is based on the Heisenberg equations of motion,
\bea\label{eomtm}
j&=&m\ddot x+m\omega^2_0x+\sum_ng_ny_n,\nn
0&=&m\ddot{y}_n+m\omega_n^2y_n+g_nx.
\eea
The elimination of the environment coordinates by their equation of motion yields
\be
y_n(\omega)=\frac{g_n}{m[(\omega+i\epsilon)^2-\omega_n^2]}x(\omega)
\ee
where the $i\epsilon$ term takes care of the initial conditions. The substitution of this equation into the first equation of \eq{eomtm} leads to the effective equation of motion, $-j(\omega)=(D^r(\omega))^{-1}x(\omega)$, in terms of the retarded Green's function,
\be\label{retgftm}
D^r(\omega)=\frac1{m[(\omega+i\epsilon)^2-\omega^2_0]-\Sigma^r(\omega)},
\ee
containing the retarded self energy,
\be
\Sigma^r(\omega)=\sum_n\frac{g^2_n}{m[(\omega+i\epsilon)^2-\omega_n^2]}.
\ee

The CTP formalism reproduces this simple result within the Lagrange formalism. The action is of the form
\be
S[\hx,\hy]=\hf\int dtdt'[\hx(t)\hD^{-1}(t,t')\hx(t')+\hy(t)\hG^{-1}(t,t')\hy(t')]-\int dt\hx(t)\sigma g\hy(t),
\ee
where $\hD(t,t')=\hG(t-t',\omega_0)/m$ and $\hG_n(t,t')=\hG(t-t',\omega_n)/m$, denote the system and environment Green's functions, respectively and the CTP matrix,
\be
\hs=\begin{pmatrix}1&0\cr0&-1\end{pmatrix},
\ee
denotes the metric tensor of the simplectic structure of the CTP formalism. A variable in a Gaussian integral can be eliminated by its classical equation of motion, so the effective action is given by 
\be
S_{eff}[\hx]=\hf\int dtdt'\hx(t)[\hD_0^{-1}(t,t')-\sigma\hat\Sigma(t,t')\sigma]\hx(t').
\ee
The self energy, $\hat\Sigma=g\hG g$, can simply be expressed in terms of the spectral function,
\be
\hat\Sigma_\omega=\frac1m\int_0^\infty d\Omega2\Omega\rho(\Omega)\hG_\omega(\Omega),
\ee
in particular
\be
\Sigma^n_\omega=\frac2mP\int_0^\infty d\Omega\frac{\Omega\rho(\Omega)}{\omega^2-\Omega^2},~~~
\Sigma^f_\omega=-i\pi\sign(\omega)\rho(|\omega|),~~~
\Sigma^i_\omega=-\pi\rho(|\omega|).
\ee
The influence functional, given by the self energy piece of the effective action, is
\be
S_{infl}[\hx]=-\hf\sum_{\sigma\sigma'}\sigma\sigma'\int\frac{d\omega}{2\pi}dte^{-i\omega(t'-t)}\Sigma^{\sigma\sigma'}_\omega x^\sigma(t')x^{\sigma'}(t).
\ee
The CTP formalism provides no more details than the traditional solution, discussed at the beginning of this section. But in the presence of interactions the CTP scheme is necessary in order to generalize Wick's theorem for open systems and to represent the contributions to the perturbation series of the Green's functions in terms of CTP Feynman graphs.

\subsection{Effective Lagrangian}
To derive the effective Lagrangian we assume that the self-energy is a real, analytic function of $i\omega$ and write the influence functional after a partial integration as
\be
S_{infl}[\hx]=-\hf\sum_{\sigma\sigma'}\sigma\sigma'\sum_{\ell=0}^\infty\frac{(-1)^\ell}{\ell!}\partial^\ell_{i\omega}\Sigma^{\sigma\sigma'}_0\int dtx^\sigma(t)\partial_t^\ell x^{\sigma'}(t).
\ee

The CTP block structure, \eq{blockg}, results in the influence Lagrangian,
\be\label{clelagr}
L_{infl}=-\hf(x\vec\Sigma^nx^d+x^d\vec\Sigma^nx+x^d\vec\Sigma^fx-x\vec\Sigma^fx^d+x^di\vec\Sigma^ix^d),
\ee
where the parametrization $x^\pm=x\pm x^d/2$ is used and the arrow on
\be\label{gradexp}
\vec\Sigma^{\sigma\sigma'}=\sum_{\ell=0}^\infty\frac{(-1)^\ell}{\ell!}\partial^\ell_{i\omega}\Sigma^{\sigma\sigma'}_0\partial_t^\ell
\ee
is a reminder that this is a differential operator and it acts to the right. The effective theory is unitary by construction and this property remains valid if the series \eq{gradexp} is truncated. This can be seen in an explicit manner by noting that the Gaussian integral, \eq{ctpeffgf}, can be found by evaluating the integrand at its saddle point, satisfying the classical equations of motion. In the case of a physical external source, $j^\sigma=\sigma j$, we have $x^+(t)=x^-(t)$, or $x^d(t)=0$, which cancels the finite, $\ord{\epsilon^0}$ part of the action \eq{ctpaction} and the influence Lagrangian, \eq{clelagr}. The resulting equation,
$W[j,-j]=0$, assures that the probability is conserved for arbitrary, physical external source.

In the case of a harmonic system, the expectation value, $\la x\ra$, satisfies the classical equation of motion and the variation equation for $x^d$ at $x^d=0$ is
\be
m\la\ddot x\ra=-(m\omega^2_0+\vec\Sigma^r)\la x\ra,
\ee
with $\vec\Sigma^r=\vec\Sigma^n+\vec\Sigma^f$. Up to order $\ord{\partial_t^2}$ we have a shift of the harmonic system potential, $\omega^2_0\to\omega^2_0-\Delta\omega^2$, with
\be
\Delta\omega^2=\frac2{m^2}\int_0^\infty d\Omega\frac{\rho(\Omega)}\Omega,
\ee
corresponding to the system potential in \eq{holagrc}, and a renormalization of the mass, $m\to m+\delta m$, where
\be
\delta m=\frac4m\int_0^\infty d\Omega\frac{\rho(\Omega)}{\Omega^3}.
\ee
Furthermore, a Newtonian friction force, $F=-k\dot x$, with $k=\pi\rho'(0)$ is generated. The imaginary part of the influence Lagrangian \eq{clelagr} yields a suppression factor, $e^{-d_0x^{d2}}$ with $d_0=\pi\rho(0)\ge0$, in the path integral \eq{otpr} and generates decoherence in the coordinate diagonal basis.

\section{Test particle in a gas}\label{partingass}
Consider now the more realistic system of a point particle of mass $m$, moving in a potential $V(\v{x})$ and interacting with a homogeneous ideal gas in equilibrium via a time-independent potential, $U(\v{x})$. The action, $S=S_p+S_g+S_i$, is the sum of the terms
\bea\label{act}
S_p[\v{x}]&=&\int dt\left[\frac{m_0}2\dot{\v{x}}^2(t)-V(\v{x}(t))\right],\nn
S_g[\psid,\psi]&=&\int dtd^3y\psid(t,\v{y})\left[i\hbar\partial_t+\frac{\hbar^2}{2m}\Delta+\mu\right]\psi(t,\v{y}),\nn
S_i[\v{x},\psid,\psi]&=&\int dtd^3yU(\v{y}-\v{x}(t))\psid(t,\v{y})\psi(t,\v{y}).
\eea

\subsection{Effective theory}\label{effths}
The influence functional,
\be\label{inflfunc}
e^{\ih S_{infl}[\hat{\v{x}}]}=\int D[\hpsi]D[\hpsid]e^{\ih S_g[\hpsid,\hpsi]+\ih S_i[\hat{\v{x}},\hpsid,\hpsi]},
\ee
contains the action of the gas particles, 
\be
S_g[\hpsid,\hpsi]+S_i[\hat{\v{x}},\hpsid,\hpsi]=\int dxdx'\hpsid(x)(\hat F^{-1}(x,x')+\hat\Gamma[x,x';\hat{\v{x}}])\hpsi(x'),
\ee
where
\be
\hat F(t,\v{y},t',\v{y}')=\int\frac{d\omega d^3q}{(2\pi)^4}e^{-i\omega(t-t')+i\v{q}(\v{y}-\v{y}')}\hat F_{\omega\v{q}},
\ee
with \cite{coulomb}
\be
\hat F_{\omega\v{q}}=\begin{pmatrix}\frac1{\omega-\omega_\v{q}+i\epsilon}&0\cr-i2\pi\delta(\omega-\omega_\v{q})&\frac1{\omega_\v{q}-\omega+i\epsilon}\end{pmatrix}-\xi n_\v{q}i2\pi\delta(\omega-\omega_\v{q})\begin{pmatrix}1&1\cr1&1\end{pmatrix},
\ee
and $\epsilon_\v{q}=\hbar\v{q}^2/2m_g$, $n_\v{q}$ denoting the occupation number. The exchange statics factor is $\xi=+1$ for bosons and $\xi=-1$ for fermions. When the Green's function is calculated in the operator representation then the excited state contributions in the initial state are picked up when the normal order, $aa^\dagger=\xi a^\dagger a+1$, is used and they generate the overall factor, $\xi n_\v{q}$, in front of the on-shell propagator of the thermal bath excitations. The temperature is supposed to be sufficiently high in the case of a bosonic ideal gas to suppress the condensate. The particle-gas interaction is described by the term $\hpsid\hat\Gamma\hpsi$ in the action, where $\Gamma^{\sigma\sigma'}[\hat{\v{x}}]=\sigma\delta^{\sigma\sigma'}\Gamma[\v{x}^\sigma]$.

The exact influence functional, corresponding to an ideal gas, 
\be
S_{infl}[\v{x}]=-i\hbar\Tr\ln[\hat F^{-1}+\hat\Gamma[\v{x}]],
\ee
is non-polynomial in the trajectory. By assuming that the test particle-gas interaction is sufficiently weak one can use $U$ to organize a perturbation expansion in which the leading, $\ord{U^2}$, order effective action is \cite{hedegard}
\be\label{infl}
S_{infl}[\hat{\v{x}}]=\frac\xi2\hj\hs\hG\hs\hj,
\ee
after canceling the $\ord{U}$ tadpoles by the introduction of a fictitious, homogeneous, neutralizing charge density. The source, $j^\sigma(t,\v{y})=U(\v{y}-\v{x}^\sigma(t))$, denotes the perturbation of the ideal gas by the test particle, and
\be
G^{\sigma_1\sigma_2}(x_1,x_2)=i\xi\hbar n_s\hat F^{\sigma_1\sigma_2}(x_1,x_2)\hat F^{\sigma_2\sigma_1}(x_2,x_1)
\ee
is the density two-point function, given in terms of the Green's functions $G^n_{\omega\v{q}}=G^+_{\omega\v{q}}+G^+_{-\omega\v{q}}$, $G^f_{\omega\v{q}}=i(G^-_{\omega\v{q}}-G^-_{-\omega\v{q}})$ and $G^i_{\omega\v{q}}=G^-_{\omega\v{q}}+G^-_{-\omega\v{q}}$, with \cite{coulomb}
\bea\label{gpm}
G^+_{\omega\v{q}}&=&-\xi\frac{n_s}{\hbar^2}P\int\frac{d^3k}{(2\pi)^3}\frac{n_\v{k}}{\omega-\omega_{\v{k}+\v{q}}+\omega_\v{k}},\nn
G^-_{\omega\v{q}}&=&-\xi\pi\frac{n_s}{\hbar^2}\int\frac{d^3k}{(2\pi)^3}n_\v{k}(n_{\v{k}+\v{q}}+\xi)\delta(\omega-\omega_{\v{k}+\v{q}}+\omega_\v{k}),
\eea
where the spin is taken into account by the degeneracy factor, $n_s=2s+1$. Since $iG^{-+}_{\omega\v{q}}=-2G^-_{\omega\v{q}}$ is the spectral function for the excited states of the gas $G^-_{\omega\v{q}}\le0$ and $G^-_{\omega\v{q}}=0$ for $\omega<0$. At finite temperatures the integrals \eq{gpm} are analytic functions of the dimensionless variables $x=\omega/|\v{q}|v_F$, $y=|\v{q}|/k_F$, $G=G(x,y)$, here $v_F=\hbar k_F/m_g$, $k_F=\sqrt[3]{6\pi^2n/n_s}$, $n$ being the density.

The $\ord{U^2}$ influence functional, \eq{infl}, is non-local in time and non-polynomial in the trajectory. A simpler structure can be recovered by resorting to the Landau-Ginsburg double expansion: We assume that the external potential, $V(\v{x})$, acting on the test particle is sufficiently attractive, allowing us to use the displacement, generated by the interactions with the gas, $|\v{x}|$, as a small parameter. The $\ord{\v{x}^2}$ approximation of the effective theory retains the leading self-energy correction to the test particle and contains dissipative forces which can be generated by a linearized equation of motion. Another small parameter is generated by restricting our attention to a slowly moving test particle where the kernel of the linearized equation of motion can be expanded in the time derivative. Such an expansion renders the effective action local in time which is expected to be the case if the gas has sufficiently high temperature. When the influence functional \eq{infl} is evaluated in Fourier space then the dominant contribution comes from the region of $\omega$ which is around the characteristic frequency of the particle motion, $\omega\sim|\v{q}\dot{\v{x}}|$, indicating that the particle should be much slower than the characteristic velocity of the gas, $|\dv{x}|\ll v_F$ and $|x|=|\omega|/|\v{q}|v_F\ll1$.

\subsection{Finite temperature}\label{quadreffs}
We are in the position to work out the effective Lagrangian, corresponding to the $\ord{\v{x}^2}$ truncation of the influence functional \eq{infl}, written as
\be\label{inflact}
S_{infl}=-\hf\sum_{\sigma\sigma'}\sigma\sigma'\int\frac{d\omega d^3q}{(2\pi)^4}dtduU_\v{q}^2e^{-i\omega u+i\v{q}(\v{x}^\sigma(t+\frac{u}2)-\v{x}^{\sigma'}(t-\frac{u}2))}G^{\sigma,\sigma'}_{\omega\v{q}}.
\ee
It yields the influence Lagrangian,
\be\label{inflbe}
L_{infl}=-\hf\sum_{\sigma\sigma'}\sigma\sigma'\int\frac{d\omega d^3q}{(2\pi)^4}duU_\v{q}^2e^{-i\omega u+i\v{q}[\v{x}^\sigma-\v{x}^{\sigma'}+\sum_{n=1}^\infty\frac{u^n}{2^nn!}(\v{x}^{(n)\sigma}-(-1)^n\v{x}^{(n)\sigma'})]}G^{\sigma,\sigma'}_{\omega\v{q}},
\ee
where the notation $\v{x}=\v{x}(t)$, $\v{x}^{(n)}=d^n\v{x}/dt^n$ is applied. We assume that the potential $V(\v{x})$ is sufficiently deep to confine the particle close to its initial position, $\v{x}=0$, and retain the leading order, $\ord{\v{x}^2}$ part only,
\be\label{inflpart}
L_{infl}=\frac14\sum_{\sigma\sigma'}\sigma\sigma'\int\frac{d^3q}{(2\pi)^3}U_\v{q}^2[\v{q}(\v{x}^\sigma-\v{x}^{\sigma'}+\Delta_+\v{x}^\sigma-\Delta_-\v{x}^{\sigma'})]^2G^{\sigma\sigma'}_{0\v{q}}.
\ee
The derivatives in
\be\label{deltax}
\Delta_\pm\v{x}=\sum_{n=1}^\infty\frac{(\pm1)^n}{2^nn!}\v{x}^{(n)}\partial_{i\omega}^n
\ee
act on the Green's function, evaluated at vanishing frequency. The block structure \eq{blockg} gives the expression
\bea
L_{infl}&=&\frac1{12\pi^2}\int_0^\infty dqq^2U_q^2\{4\Delta_1\v{x}\Delta_1\v{x}^dG^n_{0q}+4\Delta_1\v{x}(\v{x}^d+\Delta_2\v{x}^d)G^f_{0q}\nn
&&+[\Delta_1\v{x}^{d2}-(\v{x}^d-\Delta_2\v{x}^d)^2]iG^i_{0q}\}
\eea
after integrating over the direction of the wave vector $\v{q}$ and the insertion of Eq. \eq{deltax} brings the influence Lagrangian into the form
\bea
L_{infl}&=&\sum_{m,n=1}^\infty g_{2m-1,2n-1}\v{x}^{(2m-1)}\v{x}^{d(2n-1)}+\sum_{m,n=1}^\infty g_{2m-1,2n-2}\v{x}^{(2m-1)}\v{x}^{d(2n-2)}\nn
&&+i\sum_{m,n=1}^\infty d_{2m-2,2n-2}\v{x}^{d(2m-2)}\v{x}^{d(2n-2)}+i\sum_{m,n=1}^\infty d_{2m-1,2n-1}\v{x}^{(2m-1)}\v{x}^{d(2n-1)},
\eea
where the effective coupling constants,
\bea\label{cconsts}
g_{2m-1,2n-1}&=&\frac1{3\pi^22^{2(m+n-3)}(2m-1)!(2n-1)!}\int_0^\infty dqq^2U_q^2\partial_{i\omega}^{2(m+n)-2}G^+_{0q},\nn
g_{2m-1,2n-2}&=&\frac{i}{3\pi^22^{2(m+n-4)}(2m-1)!(2n-2)!}\int_0^\infty dqq^2U_q^2\partial_{i\omega}^{2(m+n)-3}G^-_{0q},\nn
d_{2m-1,2n-1}&=&\frac1{12\pi^22^{2(m+n-3)}(2m-1)!(2n-1)!}\int_0^\infty dqq^2U_q^2\partial_{i\omega}^{2(m+n)-2}G^-_{0q},\nn
d_{2m-2,2n-2}&=&\frac{(-1)^{\delta_{m,1}+\delta_{n,1}+1}}{3\pi^22^{2(m+n-3)}(2m-2)!(2n-2)!}\int_0^\infty dqq^2U_q^2\partial_{i\omega}^{2(m+n)-4}G^-_{0q}
\eea
have been introduced. The real terms in the Lagrangian contribute to the equation of motion and the even and odd order derivative terms, generated by $G^n$ and $G^f$, represent time-reversal invariant interactions and friction forces, respectively. The imaginary part of the effective Lagrangian originates from $G^i$ and governs the decoherence. The constants \eq{cconsts} introduce the time scales $\tau_{m,n}=(|g_{m,n}|/m)^{1/(m+n-2)}$ and $\tau'_{m,n}=(|d_{m,n}|/m)^{1/(m+n-2)}$ with $m+n\ne2$ in the effective dynamics. The time-reversal parity of the term in the Lagrangian which is multiplied by $g_{m,n}$ is $(-1)^{m+n}$ therefore $\tau_{m,n}$ is an intrinsic time scale of the conservative or the dissipative effects for $n+m$ even or odd, respectively. The characteristic time scales of the decoherence, $\tau'_{m,n}$, are generated by time-reversal invariant terms.

It is important to realize that both the irreversibility and the decoherence stem from the same source, namely from the Green's function component $G^-$. If the time arrow of the gas is flipped then $G^-$ changes sign, indicating that $G^-$ represents the dynamical breakdown of the time-reversal symmetry, the dynamical origin of dissipative processes \cite{breakd}. Therefore dissipation appears with different time scales in the effective dynamics, some of them corresponding to irreversibility, others belonging to decoherence. The common origin of irreversibility and decoherence can clearer be seen by noting that irreversibility has a genuine quantum manifestation, the leakage of the quantum state into the environment. This is characterized by the life-time of the state, given by the imaginary part of the inverse Feynman propagator in the frequency space. The self-energy of the test particle is given by the contributions $\sigma=\sigma'$ on the right hand side of Eq. \eq{inflpart} and the life-time arises from $\Im G^{\pm\pm}$. The decoherence is driven by the imaginary part of the terms $\sigma=-\sigma'$, containing $\Im G^{\pm\mp}$. The Green's function $G^{\sigma\sigma'}$ possesses the block structure \eq{blockg} hence both the life-time and the decoherence are driven by the same Green's function component, namely by $G^i$; in particular they have the same characteristic time scale.

The higher order derivatives in the equation of motion create difficulties because the corresponding initial conditions are unknown \cite{breakd}. Hence we restrict ourselves to the order $\ord{\partial_\omega^2}$ where the effective Lagrangian,
\be\label{linflk}
L_{eff}=(m_0+\delta m)\dv{x}^d\dv{x}-k\v{x}^d\dv{x}+\frac{i}2[d_0(\v{x}^d)^2+d_2(\dv{x}^d)^2],
\ee
contains the parameters 
\be\label{emcoeff}
\delta m=g_{1,1},~~~k=-g_{1,0},~~~d_0=d_{0,0},~~~d_2=d_{1,1}
\ee
One can prove that the dynamics, induced by such a truncated Lagrangian, is unitary by repeating the argument, mentioned above in relation to the Lagrangian \eq{clelagr}.

The Euler-Lagrange equation for $\v{x}^d$ at $\la\v{x}^d\ra=0$,
\be\label{relpingem}
m\la\ddv{x}\ra=-\la\v{\nabla}V(\v{x})\ra-k\la\dv{x}\ra+\ord{\partial_t^3}+\ord{\la\v{x}^3\ra},
\ee
given by $\Re L_{infl}$, shows the mass renormalization, $m=m_0+\delta m$, and the presence of a non-vanishing Newtonian friction force. Note that the $\ord{k}$ friction term generates changes of the momentum by the time without violating the translation invariance of the influence functional. In the calculation of ref. \cite{hedegard} $G^f$ was used in the place of $G^n$ and so no mass renormalization is found. The dissipative term of the equation of motion \eq{relpingem} has been found also by the collisional scheme, the friction constant being expressed in terms of the $T$ matrix \cite{vacchinie}.

\subsection{Electron gas at vanishing temperature}\label{idfgvts}
If the temperature approaches zero in a fermionic ideal gas then the Green's function, $\hG_{\omega\v{q}}$, simplifies and develops singularities. The loop-integrals in eqs. \eq{gpm} can easily be calculated for $n_\v{q}=\Theta(k_F-|\v{q}|)$ \cite{coulomb} and $G^+$, the Lindhard function, is found to be analytical for $|x|+y/2<1$. Within this domain we have  $G^-=-k_Fm_g\Theta(x)x/2\pi\hbar^2$, yielding $G^f=-ik_Fm_gx/2\pi\hbar^2$ and $G^i=-k_Fm_g|x|/2\pi\hbar^2$. The influence Lagrangian, using the same approximation as before, is now of the form
\be\label{ztrl}
\Re L_{infl}=-k\v{x}^d\dv{x}+\delta m\dv{x}^d\dv{x}+\ord{\partial_t^4}+\ord{\v{x}^4},
\ee
with the friction constant
\be\label{frczt}
k=\frac{m^2_g}{12\pi^3\hbar^3}\int_0^{2k_F}dqq^3U_q^2,
\ee
and $\delta m$, given by \eq{emcoeff}. Observe that there are no higher order irreversible terms owing to $D^f(x,y)\sim x$. 

In calculating $\Im L_{infl}$ the expansion of the exponent on the right-hand side of Eq. \eq{inflbe} must be carried out around $iu(\v{q}\dot{\v{x}}-\omega)$, yielding
\be
\Im L_{infl}=\frac14\sum_{\sigma\sigma'}\sigma\sigma'\int\frac{d^3q}{(2\pi)^3}U_\v{q}^2[\v{q}(\v{x}^\sigma-\v{x}^{\sigma'})]^2\Im G^{\sigma\sigma'}_{\v{q}\dv{x}^\sigma,\v{q}}.
\ee
Note that $\Delta\v{x}$, which now starts at the order $\ord{\partial_{i\omega}^2}$, can be ignored due to $G^i\sim|x|$. The block structure, \eq{blockg}, gives
\be
\Im L_{infl}=-\frac14\int\frac{d^3q}{(2\pi)^3}U_\v{q}^2(\v{q}\v{x}^d)^2\left[G^i_{\v{q}(\dv{x}+\frac{\dv{x}^d}2),\v{q}}+G^i_{\v{q}(\dv{x}-\frac{\dv{x}^d}2),\v{q}}\right]
\ee
which can be replaced by
\be\label{ztil}
\Im L_{infl}=\frac{k_Fm_g}{4\pi\hbar^2}\int\frac{d^3q}{(2\pi)^3}U_q^2G^i_{\v{q}\dv{x},\v{q}}(\v{q}\v{x}^d)^2
\ee
when the contributions beyond $\ord{\v{x}^2}$ are neglected. The integration over the solid angle is straightforward and leads to a particular anisotropy,
\be\label{ldec}
\Im L_{infl}=i\lambda|\dv{x}|(\v{x}_t^{d2}+6\v{x}_\ell^d)
\ee
where $\v{x}^d$ was separated into longitudinal and transverse components, $\v{x}^d_\ell$, $\v{x}^d_t$, respectively, defined by $\v{x}^d=\v{x}^d_\ell+\v{x}^d_t$, $\v{x}^d_t\dv{x}=0$ and
\be\label{lam}
\lambda=\frac{m^2_g}{64\pi^3\hbar^3}\int_0^{2k_F}dqq^4U_q^2.
\ee

It is easy to understand the origin of the factor $|\dv{x}|$ in \eq{ldec}. The dissipative processes and the decoherence are generated by an energy exchange with the environment and this is always possible if the gas is in thermal equilibrium. For vanishing temperature the gas is in its ground state and the particle must possess some kinetic energy to interact with the gas. This condition is automatically satisfied by the friction force, being proportional to the velocity. But the decoherence would start in zeroth order if the particle is coupled to the gas by its density. The role of the factor $|\dv{x}|$ is to suppress the system-environment correlations for a particle at rest. The initial state of the toy model is not at the energy minimum which explains that the decoherence is present even for a static system. 

The absence of further, higher order derivative dissipative forces and of the decoherence for a particle at rest remains valid when the higher powers of the coordinate are retained in the effective Lagrangian as long as the gas is eliminated in the one-loop approximation, i.e. Eq. \eq{infl} applies.

\section{Positivity of the density matrix}\label{masters}
When an approximation is applied one has always to check whether the consistency conditions for the solution remain valid. The effective theory for the test particle describes the reduced density matrix which has to satisfy a number of conditions, such as the preservation of the total probability, the Hermiticity, and the positivity. It has already been mentioned at the end of Sec. \ref{shos} that the block structure \eq{blockg} makes the total probability preserved for a harmonic oscillator. It is easy to check that the same block structure of the Green's function is recovered when calculated in the one-loop approximation and it guarantees the Hermiticity of the density matrix, too. Thus what is left to check is the positivity. This is easier to do by inspecting the equation of motion for the density matrix, generated by the effective Lagrangian \eq{linflk}.

\subsection{Master equation}

The nowhere differentiable nature of the trajectories, dominating the path integrals, makes it important to use the mid-point prescription in the case of  coordinate and velocity couplings \cite{schulmann}. This applies to the $\ord{k}$ friction term of the Lagrangian which is generated by $G^f(-\omega)=-G^f(\omega)$ and therefore determined by the symmetric time derivative, $-k\v{x}^d_n(\v{x}_{n+1}-\v{x}_{n-1})/2\dt$ for discrete time steps where $f_n=f(t_i+n\dt)$. The expression $\v{x}_n^d(\v{x}_{n+1}-\v{x}_{n-1})$, written as $\v{x}_n^d(\v{x}_{n+1}-\v{x}_n)-\v{x}_{n-1}(\v{x}^d_n-\v{x}^d_{n-1})+\v{x}_n\v{x}^d_n-\v{x}_{n-1}\v{x}^d_{n-1}$, can be replaced by $\v{x}_n^d\v{x}_{n+1}-\v{x}_n\v{x}^d_{n+1}+\v{x}_n\v{x}^d_n-\v{x}_{n-1}\v{x}^d_{n-1}$ in the coupling of two subsequent time slices in the path integral. This step leads to the regularized Lagrangian,
\bea\label{discrl}
L_B(\hat{\v{x}}_{n+1},\hat{\v{x}}_n)&=&m\frac{(\v{x}_{n+1}-\v{x}_n)(\v{x}_{n+1}^d-\v{x}_n^d)}{\dt^2}-\frac{k}2\frac{\v{x}_n^d\v{x}_{n+1}-\v{x}_n\v{x}_{n+1}^d+\v{x}_{n+1}\v{x}^d_{n+1}-\v{x}_n\v{x}^d_n}\dt\nn
&&-V(\v{x}_{n+1}^+)+V(\v{x}_{n+1}^-)+\frac{i}2\left[d_0(\v{x}_{n+1}^d)^2+d_2\left(\frac{\v{x}_{n+1}^d-\v{x}_n^d}{\dt}\right)^2\right].
\eea

We first identify the bare Liouville operator, ${\cal U}_B$, leading the time evolution over a discrete time step $\dt$. Its action on the density matrix can be written in the path integral formalism as
\be\label{ndensm}
{\cal U}_B\rho(\hat{\v{x}})=\left(\frac{m_0}{2\pi\dt\hbar}\right)^3\int d^3yd^3y^de^{\ih\dt L_B(\hat{\v{x}},\hat{\v{x}}+\hat{\v{y}})}\rho(\hat{\v{x}}+\hat{\v{y}}),
\ee
the normalization being given in terms of the bare mass, $m_0$. Working up to $\ord{\dt}$ one finds
\bea\label{infevdm}
{\cal U}_B\rho(\hat{\v{x}})&=&\left(\frac{m_B}{2\pi\dt\hbar}\right)^3e^{-\ih\dt V(\v{x}^+)+\ih\dt V(\v{x}^-)-\frac{d_0\dt}{2\hbar}\v{x}^{d2}}\nn
&&\times\int d^3yd^3y^de^{\frac{im}{\hbar\dt}\v{y}\v{y}^d+\frac{ik}{2\hbar}(2\v{x}^d\v{y}+\v{y}\v{y}^d)-\frac{d_2}{2\hbar\dt}\v{y}^{d2}}\left[1+\hy\hat{\v{\nabla}}+\hf(\hy\hat{\v{\nabla}})^2\right]\rho(\hat{\v{x}},t).
\eea
It is easy to check that any other distribution of $V(\v{x}^\sigma)$ and $\v{x}^{d2}$ between the two consecutive time steps in the Lagrangian \eq{discrl} yields the same results in $\ord{\dt}$. The Gaussian integration in Eq. \eq{infevdm} leads to the differential equation
\be\label{neumang}
\partial_t\rho=\frac1{i\hbar}[H,\rho]+{\cal D}\rho,
\ee
with $H=\v{p}^2/2m+V(\v{x})$, the role of the generator ${\cal D}$ being played by 
\be
{\cal D}_B=-\frac{\tilde d}{2\hbar}\v{x}^{d2}-\frac{k}m\v{x}^d\v{\nabla}_{x^d}+\frac{id_2k}{m^2}\v{x}^d\v{\nabla}_x+\frac{d_2\hbar}{2m^2}\Delta_x+\frac3\dt\ln\frac{m_0}{m},
\ee
and $\tilde d=d_0+d_2k^2/m^2$ denoting an effective decoherence strength parameter. 

Let us now assume, for the sake of the argument, that the particle moves in a harmonic potential, $V(\v{x})=m\omega^2\v{x}^2/2$, and its action can be recast in the form \eq{freectpact}. Since the Green's function, $\hG$, appearing in the influence functional \eq{infl}, possesses the block structure of Eq. \eq{blockg}, the generator functional \eq{ctpgfunc} of the particle is vanishing for physically realizable sources, $j^+=-j^-$, even if it is interacting with the gas within our approximation. This result indicates that the total probability is preserved in our calculation as can be realized by the renormalization,
\be\label{normgen}
{\cal D}={\cal D}_B-\frac3\dt\ln\frac{m_0}{m},
\ee
which takes into account the trajectory independent contributions to the effective Lagrangian, neglected in the derivation. Since these are independent of $V(\v{x})$, the result remains valid for arbitrary $V(\v{x})$. The action of the renormalized generator on the density matrix can be written in a more useful form,
\be\label{ngenrb}
{\cal D}\rho=\frac{ik}{2m\hbar}[\{\v{p},\rho\},\v{x}]+\frac{d_2}{2m^2\hbar}[[\v{p},\rho],\v{p}]+\frac{\tilde d}{2\hbar}[[\v{x},\rho],\v{x}]+\frac{d_2k}{m^2\hbar}[[\v{p},\rho],\v{x}].
\ee
The master equation, derived within the $T$-matrix formalism \cite{diosi,vacchini} is of this form except that it is lacking the last term. The latter represents an interference between the friction and the $\ord{\partial_t^2}$ decoherence terms of the effective Lagrangian \eq{linflk} and is not explicitly present in the collision based approach.

The equations of motion originating from the quadratic Lagrangian \eq{linflk} are satisfied by the expectation value of the coordinate, $m\la\ddv{x}\ra=-m\omega^2\la \v{x}\ra-k\la\dv{x}\ra$. It is easy to find the equation of motion for the expectation values of the canonical operators for an arbitrary potential, $V(\v{x})$,
\bea\label{deqm}
m\la\dot{\v{x}}\ra&=&\la\v{p}\ra\nn
\la\dot{\v{p}}\ra&=&-\la\v{\nabla}V(\v{x})\ra-\frac{k}{m}\la\v{p}\ra,
\eea
which lead to the Euler-Lagrange equation of the effective Lagrangian \eq{linflk},
\be\label{eleom}
m\la\ddv{x}\ra=-\la\v{\nabla}V(\v{x})\ra+k\la\dv{x}\ra.
\ee

\subsection{Positivity}
The density matrix preserves its positivity as long as the the master equation can be written in the Lindblad form \cite{gerhardts,lindblad},
\be
\dot\rho=\frac1{i\hbar}[H_L,\rho]+\frac1{2\hbar}\sum_j\left([V_j\rho,V_j^\dagger]+[V_j,\rho V_j^\dagger]\right).
\ee
The generator \eq{ngenrb} of the time evolution can be written in such a form with the help of the operator set $\{\v{x},\v{p},\v{a},\v{b}\}$, where
\be
\v{a}=\frac{m\bar\omega_0\v{x}+i\v{p}}{\sqrt{2m\hbar|\bar\omega_0|}},~~~
\v{b}=\frac{m\omega_0\v{x}+\v{p}}{\sqrt{2m\hbar|\bar\omega_0|}},
\ee
$\omega_0$ standing for an arbitrary frequency,
\bea\label{lindbf}
{\cal D}\rho&=&\frac1{i\hbar}\left[\Delta H,\rho\right]+D_x([\v{x}\rho,\v{x}]+[\v{x},\rho\v{x}])+D_p([\v{p}\rho,\v{p}]+[\v{p},\rho\v{p}])\nn
&&+D_a([\v{a}\rho,\v{a}^\dagger]+[\v{a},\rho\v{a}^\dagger])+D_b(\left[\v{b}\rho,\v{b}\right]+\left[\v{b},\rho\v{b}\right]),
\eea
with
\bea\label{lindblp}
\Delta H&=&\frac{k}{4m}\{\v{x},\v{p}\}\nn
\hbar D_x&=&\frac{\tilde d}2-\frac{\bar\omega k}2\left(\hf+\frac{d_2}m\right),\nn
\hbar D_p&=&\frac1{2m^2}\left[d_2-\frac{k}{\bar\omega}\left(\hf+\frac{d_2}m\right)\right],\nn
D_a&=&\sign(\bar\omega)\frac{k}{2m},\nn
D_b&=&\sign(\bar\omega)\frac{d_2k}{m^2}.
\eea
The resulting equation is a special case of the possible Lindblad-compatible  master equations which are bilinear in the canonical variables \cite{sandulescu}. To assure $D_x,D_p>0$, we have to impose the inequality
\be
k\left(\frac1{2d_2}+\frac1m\right)<\bar\omega<\frac{2m\tilde d}{k(m+2d_2)},
\ee
which is always possible as long as
\be\label{lindblineq}
\nu<2\sqrt{\frac{d_0d_2}{m(m+4d_2)}}.
\ee
Note that both decoherence parameters up to $\ord{\partial_t^2}$, $d_0$ and $d_2$, are needed to establish friction and one needs a finite temperature for an ideal fermion gas to generate the positive density matrix in our approximation.

We may generate the $\ord{k}$ surface term in the Lagrangian \eq{discrl} by a time independent basis (gauge) transformation, $\psi(\v{x})\to{\cal G}\psi(\v{x})=e^{\frac{ik}{4\hbar}\v{x}^2}\psi(\v{x})$, and use the Lagrangian,
\bea
L'_B(\hat{\v{x}}_{n+1},\hat{\v{x}}_n)&=&m\frac{(\v{x}_{n+1}-\v{x}_n)(\v{x}_{n+1}^d-\v{x}_n^d)}{\dt^2}-\frac{k}2\frac{\v{x}_n^d\v{x}_{n+1}-\v{x}_n\v{x}_{n+1}^d}\dt\nn
&&-V(\v{x}_{n+1}^+)+V(\v{x}_{n+1}^-)+\frac{i}2\left[d_0(\v{x}_{n+1}^d)^2+d_2\left(\frac{\v{x}_{n+1}^d-\v{x}_n^d}{\dt}\right)^2\right].
\eea
The calculation followed above gives the generator
\bea\label{sdgen}
{\cal D}'\rho&=&\frac{ik}{2m\hbar}(\v{p}\rho\v{x}-\v{x}\rho\v{p})+\frac{ik^2}{8m\hbar}(\v{x}^2\rho-\rho\v{x}^2)-\frac{d_2}{2m^2\hbar}(\v{p}^2\rho+\rho\v{p}^2-2\v{p}\rho\v{p})\nn
&&-\frac{\tilde d'}{2\hbar}(\v{x}^2\rho+\rho\v{x}^2-2\v{x}\rho\v{x})-\frac{d_2k}{2m^2\hbar}(\v{x}\v{p}\rho-\v{x}\rho\v{p}-\v{p}\rho\v{x}+\rho\v{p}\v{x})+\frac{3k}{2m}\rho,
\eea
with $\tilde d'=d_0+d_2k^2/4m^2$ which can be brought into the Lindblad form \eq{lindbf} by
\bea
\Delta H&=&-\frac{k^2}{8m}\v{x}^2\nn
\hbar D_x&=&\frac{\tilde d'}2-\frac{\bar\omega k}4\left(1+\frac{d_2}m\right),\nn
\hbar D_p&=&\frac1{2m^2}\left[d_2-\frac{k}{2\bar\omega}\left(1+\frac{d_2}m\right)\right],\nn
D_a&=&\sign(\bar\omega)\frac{k}{2m},\nn
D_b&=&\sign(\bar\omega)\frac{d_2k}{2m^2}.
\eea
The coefficients can be chosen positive if
\be
\nu<2\sqrt{\frac{d_0d_2}{m(m+2d_2)}},
\ee
and the corresponding equations of motion are
\bea\label{seqm}
m\la\dot{\v{x}}\ra&=&\la\v{p}\ra-\frac{k}2\la\v{x}\ra\nn
\la\dot{\v{p}}\ra&=&-\la\v{\nabla}V(\v{x})\ra-\frac{k}{2m}\la\v{p}\ra+\frac{k^2}{4m}\la\v{x}\ra.
\eea
If $\v{p}$ is eliminated then the equation of motion for $\v{x}$ agrees with Eq. \eq{eleom}. Notice, however, that the gauge transformation $\cal G$ leads to the representation $\v{p}\to\v{p}+k\v{x}/2$, which triggers the breaking of translation invariance by $\Delta H$.

\section{Conclusions}\label{concls}
A simple scheme was presented to calculate the effective Lagrangian of a particle which performs a slow, small amplitude motion in a static potential and interacts with an ideal gas. The calculation, performed on the one-loop level, is based on the Landau-Ginsburg double expansion in the particle trajectory and the time derivative. Irreversibility appears in two different manners in the effective theory: The friction arises from the time-reversal odd terms of the real part of the effective Lagrangian and the decoherence originates from the imaginary part. A Newtonian friction force and a mass renormalization were found in the linearized equation of motion, considered in the $\ord{\partial_t^2}$ order. Furthermore the decoherence of the particle coordinate has been established up to the same order. In the case of a Fermi gas at vanishing temperature the higher derivative dissipative forces and the decoherence of the particle momentum are absent in the one-loop effective dynamics. 

It is instructive to compare the dissipative forces with the Abraham-Lorentz force, the ``friction force'' of the electromagnetic radiation field. The Newtonian friction force, discussed in this work in the context of the electron gas can be considered as the radiation reaction force, generated by the emission of particle-hole pairs. This friction force allows the measurement of the velocity with respect to the environment and it appears if the system looses either the Galilean- or the Lorentz-boost invariance due to the coupling to its environment or due to the initial conditions. If the electromagnetic field is boost invariant in the initial state then the diffusive part of the radiation reaction force must contain at least the third time derivative of the charge trajectory.

In the case of a finite temperature fermion gas and a sufficiently weak friction the master equation, generated by our effective Lagrangian can be brought in the Lindblad form. It is remarkable that both the full $\ord{\partial_t^2}$ Lagrangian and a finite temperature environment are needed to establish this result.

Unfortunately, a detailed comparison of the coefficients occurring in the master equations obtained by different approaches is highly non-trivial due to the different assumptions involved. Although the master equation obtained in the Born approximation shows some superficial similarity with Eq. \eq{emcoeff}, important differences exist because the latter is not based on the scattering picture. The derivation of the master equation within the harmonic toy model is usually based on the spectral representation of the environment oscillators. Our master equation \eq{ngenrb} follows as soon as the influence Lagrangian \eq{clelagr} is truncated at $\ord{\partial_t^2}$, yielding the effective Lagrangian \eq{linflk}.

The higher order terms of the Landau-Ginsburg expansion can be included in our procedure, while this task represents a real challenge in the collisional approach. A more careful treatment of the non-Markovian memory terms arising naturally in our scheme is possible, and it should provide a non-trivial, physically motivated example of non-Markovian behavior \cite{breuer}.

\end{document}